\documentclass[twocolumn,prl,aps,showpacs]{revtex4}

\usepackage{graphicx}%
\usepackage{dcolumn}

\newcommand{\als}{\alpha_s}
\newcommand{\beq}{\begin{equation}}
\newcommand{\eeq}{\end{equation}}
\newcommand{\bea}{\begin{eqnarray}}
\newcommand{\eea}{\end{eqnarray}}
\newcommand{\as}{a_s}

\begin{document}    

\title{
\vskip-0.1cm

{\baselineskip 11pt
\centerline{\normalsize{}\hfill THEP 01/13}
\centerline{\normalsize\hfill TTP01-19 }
\centerline{\normalsize\hfill hep-ph/0108197}
\centerline{\normalsize\hfill August 2001}
{}}
\vskip.4cm

The Cross Section of $e^+ e^-$ Annihilation into Hadrons of Order  
$\alpha_s^4 n_f^2$ \\
in Perturbative QCD
\vskip.3cm
     }
\author{P.~A.~Baikov}
\affiliation{Institute of Nuclear Physics, Moscow State University,
Moscow~119899, Russia
        }
\author{K.~G.~Chetyrkin}\thanks{{\small Permanent address:
Institute for Nuclear Research, Russian Academy of Sciences,
 Moscow 117312, Russia}.}
\affiliation{ Fakult{\"a}t f{\"u}r Physik
 Albert-Ludwigs-Universit{\"a}t Freiburg,
D-79104 Freiburg, Germany }
\author{J.~H.~K\"uhn}
\affiliation{Institut f\"ur Theoretische Teilchenphysik,
  Universit\"at Karlsruhe, D-76128 Karlsruhe, Germany}

\begin{abstract}
\vspace*{.7cm}
\noindent
We present  the first genuine QCD five-loop calculation
of the vacuum  polarization functions: analytical  
terms of order $\alpha_s^4 n_f^2$ to the
absorptive parts of vector and scalar correlators.  
These corrections  form  an important   gauge-invariant 
subset of the full ${\cal O}(\als^4)$ correction to
$e^+ e^-$ annihilation  into hadrons and the Higgs decay rate
into hadrons respectively. They discriminate between different
widely used estimates of  the full result. 
\end{abstract}


\maketitle

The value of the strong coupling $\als$ is one of the fundamental
constants of Nature, comparable in its importance to the
electromagnetic fine structure constant $\alpha_{\rm em}$ or the weak
coupling $\alpha_{\rm weak}$, characterizing the coupling strength of
the charged $W$ boson to lefthanded fermions.  A precise measurement
of $\als$ is important for the prediction of numerous observables in
the framework of Quantum Chromodynamics (QCD).  The comparison between
the values determined with the methods of perturbative QCD (pQCD) at high
energies and from lattice simulations  respectively gives insight in
to the quality of approximation methods and the validity of the
overall framework. The combination of precise values of the three
couplings $\als$, $\alpha_{\rm em}$ and $\alpha_{\rm weak}$ in the
context of Grand Unified Theories leads, last but not least,
to important information on the structure and the particle content of
these theories beyond the Standard Model.

From the theoretical viewpoint the cross section for electron-positron
annihilation into hadrons (or its variants like the $Z$-boson decay
rate into hadrons or the semileptonic branching ratio of the
$\tau$-lepton into hadrons) can be reliably predicted in pQCD and thus
leads to one of the ``gold-plated'' $\als$ determinations.

Nontrivial QCD effects in these observables start at order $\als^2$
and their early evaluation \cite{CheKatTka79+NPB80} has laid the
ground for the subsequent developments, both on the experimental and
theoretical side (for a review see, e.g. \cite{ckk96}).  During the past
years, in particular through the analysis of $Z$ decays at LEP and of
$\tau$ decays, an enormous reduction of the experimental uncertainty
(down to 3\%) has been achieved with the perspective of a further
reduction by a factor of four at a future linear collider \cite{TESLA}.

Inclusion of the ${\cal O}(\als^3)$ corrections \cite{GorKatLar91SurSam91} 
is thus mandatory
already now. Quark mass effects as well as corrections specific to the
axial current \cite{KniKue90b} must be included for the case of
$Z$-decays.  The remaining theoretical uncertainty from uncalculated
higher orders is at present comparable to the experimental one
\cite{QCDreview}. This has led to numerous estimates of the ${\cal
O}(\als^4)$ terms
based on principles like that of ``minimal sensitivity'' (PMS), 
or of fastest apparent convergence (FAC) or by
invoking the large $\beta_0$ limit (``naive nonabelization'' (NNA))
etc. 
\cite{ste,gru,BG}. In fact, already now the
extraction of $\alpha_s$ from $\tau$-decays depends crucially on the
corresponding estimates (see, e.g. \cite{QCDreview}).

The new calculational methods used in \cite{CheKatTka79+NPB80}, in
particular recurrence relations based on the method of integration by
parts \cite{me81b}, had opened the door also to the four-loop
calculation. Nevertheless, it took more than ten years of refinement
of theoretical tools and development of the algebraic programs to move
from the three to the four-loop calculation \cite{GorKatLar91SurSam91}.

It should be noted that much of this effort went into development of
the algebraic program MINCER \cite{mincer2} written in FORM \cite{Ver91},
which implements the integration-by-parts algorithm to evaluate
massless three loop propagators. An important part of the calculation,
the generation  of diagrams and their ultraviolet renormalization,
was performed manually. In this context it is essential that one
initially encounters the absorptive parts of four-loop amplitudes
whose evaluation is reduced to that of three-loop propagators,
employing the $R^*$ operation \cite{me84}.   That
latter step was also performed manually --- an error-prone and
time-consuming procedure difficult to implement in computer algebra,
hence presenting a real obstacle for the ${\cal O}(\als^4)$
calculation.
 
Recently the order ${\cal O}(\als^3)$ result has been rederived in a
general covariant gauge in a completely automatic way
\cite{gvvq}. This latter work was based on a general formula which
explicitly expresses the result in terms of well-known renormalization
constants and three-loop massless propagators, thus resolving the
complicated combinatorics of ultraviolet and infrared subtractions.
This formula is valid for any number of loops and is, therefore, an
important ingredient of the present ${\cal O}(\als^4)$ calculation.

The calculational
effort for a full evaluation of $R(s)$ in ${\cal O}(\als^4)$ is
enormous and exceeds the presently available computer resources by
perhaps two orders of magnitude.  It is for this reason that we limit
the present calculation to a gauge invariant subset, namely the terms
of order $\als^4 n_f^2$, where $n_f$ denotes the number of fermion
flavours.  The terms of order $\als^4 n_f^3$ (and, in fact, all terms
of order $\als(\als n_f)^n$) have been obtained earlier by summing the
renormalon chains \cite{VV:renormalons}. These give interesting
insight into the behavior of perturbation series in high orders and
have been confirmed by the present work. However, they are numerically
small   and not directly sensitive to the
nonabelian character of QCD, in contrast to the terms of order $\als^4
n_f^2$.

The motivation of the present work is threefold. First: presenting the
first ${\cal O}(\als^4)$ results in pQCD  demonstrates that the
methods proposed in \cite{gvvq,bai1,bai2} are indeed suited to obtain
genuine QCD results in five loop approximation. Second: the
results per se are an important ingredient of $R(s)$. And third: when
comparing the exact $\als^4 n_f^2$ results with the estimates based on
various optimization procedures one obtains important insights into
the quality of different approaches, confirming some and refuting
others.

Below we will present the results for the vector correlator and the
corresponding quantity $R(s)$ discussed above, and for the
semileptonic branching ratio of the $\tau$~lepton. The same methods
are applied for the scalar correlator. In this case the results are of
relevance for the Higgs decays into quarks, for the QCD sum rule
determinations of the strange quark mass and, last not least, they
provide a second independent check of the ``optimization procedures''.

Let us now briefly describe the methods used for the practical
calculations. 
Using the techniques from \cite{gvvq} and the known renormalization
constants up to order $\als^4$ the four loop propagator type massless
integrals remain to be evaluated. They were calculated following the
method described in \cite{bai1,bai2}. It is based on the integration
by part (IBP) technique \cite{me81b}, however with important
modifications. The integral is to be reduced to a sum of irreducible
(``master'') integrals with coefficients that are known to be rational
functions of the space-time dimension $D$. However, in contrast to the
original approach, where these coefficients were calculated by a
recursive procedure, they are in the current approach obtained from an
auxiliary integral representation\cite{bai1}.

For three-loop vacuum integrals these representations can be solved
explicitly in terms of Pochhammers symbols \cite{bai1}.  The structure
of four-loop propagators is much more involved and an explicit
solution is not available. In the present case the following 
method  is
used: for massless propagators the coefficients of the master
integrals should be rational functions of space-time dimension $D$
only. Their integral representation is expanded in the limit
$1/D\rightarrow 0$. Calculating sufficiently many terms in this
expansion, the original $D$-dependence can be reconstructed. This
approach is essentially different from $1/D$ expansion for the {\em
very} Feynman integrals, recently proposed in Ref.~\cite{Tar}.  
In our case only the coefficients of the master integrals (rational
functions in $D$!) are expanded and the original integral can be
reconstructed exactly.

In practice we proceed as follows: First, the set of irreducible
integrals involved in the problem was constructed, using the criterion
irreducibility of Feynman integrals \cite{bai2}.  Second, the
coefficients multiplying  these master integrals were calculated in the
$1/D\rightarrow0$ expansion. This  part was
performed using the parallel version of FORM \cite{Fliegner:2000uy} running on an
8-alpha-processor-SMP-machine with disk space of 350 GB.  The
calculations took approximately 500 hours in total. (The
time includes $\approx$ 300 hours spent on the single diagram
presented on Fig.~1a, which demonstrates that trivial parallelization
by assigning different diagrams to different machines is not
feasible.)  Third, the exact answer was reconstructed from results of
these expansions.  Extensive tests were performed.  
Dealing with four-loop integrals, the result cannot have the
singularities stronger then $1/(D-4)^4$ in $D\rightarrow4$ limit. The
contributions corresponding to individual master integrals have poles
up to $1/(D-4)^6$ which  cancel in the final
expression. Furthermore, singular parts of all diagrams were checked
by MINCER.  Previously obtained four-loop results were successfully
reproduced.
\begin{figure}[h]
\includegraphics{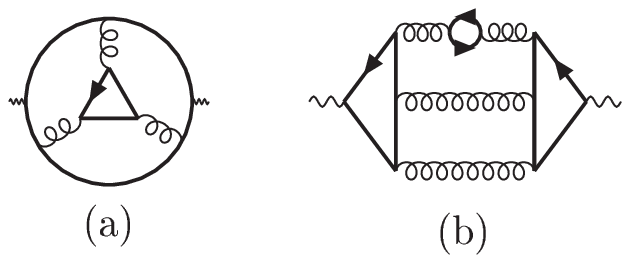}
\caption{(a) Diagram whose calculation took $\approx$ 300 hours. (b)
Diagram contributing to the QED $\beta$-function in
${\cal O}(\alpha_{\rm em}^6 n_f^3)$.}
\end{figure}

It is convenient to define ``Adler functions'' as
\beq
\nonumber
D(Q^2) =  \int_0^\infty \frac{Q^2\ R(s) d s }{(s+Q^2)^2}
{}, \  \ 
\widetilde{D}(Q^2) =  \int_0^\infty \frac{ Q^2\ \widetilde{R}(s) d s }{(s+Q^2)^2}
{}\ .
\eeq 
Here, $R(s) = 1+ \alpha_s/\pi + \dots$ is the famous $R$-ratio in the
massless limit (we suppress the trivial factor $\sum_f Q^2_f$
throughout) and $\widetilde{R}(s) = 1+ \frac{17}{3} \alpha_s/\pi +
\dots $ is properly normalized imaginary part of the scalar
polarization function. 
Our results for $D, \widetilde{D}, R$ and , $\tilde{R}$ are presented
below. Let us define ($\as \equiv \als/\pi$)
\bea
 D(Q^2) &=& 1+ \sum_{i=1}^{\infty} d_i\  a_s^i(Q^2), \ \ 
R(s)   =  1 + \sum_{i=1}^{\infty} r_i \ a_s^i(s)\ ,
\nonumber
\\ 
\nonumber
 \widetilde{D}(Q^2) &=& 1+ \sum_{i=1}^{\infty} \  \widetilde{d}_i a_s^i(Q),  \ \ 
\tilde{R}(s) = 1 + \sum_{i=1}^{\infty} \  \tilde{r}_i a_s^i(s)
{}\ ,
\eea
where we have set the normalization scale $\mu^2=Q^2$ or to $\mu^2= s$
for the euclidian and minkowskian functions respectively.  The results
for generic values of $\mu$ can be easily recovered with the standard
RG techniques. Since the functions $D$ and $\tilde{D}$
are known to order $\als^3$ from
Refs.~\cite{GorKatLar91SurSam91,gvvq,gssq} we cite below only the 
${\cal O}(\als^4)$ results 
(dots stand for still unknown terms of order $\als^4 n_f$ and $\als^4 n_f^0$).
\begin{eqnarray}
&&d_4 =   
 \,C_F   \, n_f^3
\left[
-\frac{6131}{7776} 
+\frac{203}{432}  \,\zeta_{3}
+\frac{5}{24}  \,\zeta_{5}
\right]
\nonumber\\
&{+}& \,C_F \,C_A   \, n_f^2
\left[
\frac{340843}{20736} 
-\frac{10453}{1152}  \,\zeta_{3}
-\frac{1}{8}  \,\zeta_3^2
-\frac{85}{18}  \,\zeta_{5}
\right]
\nonumber\\
&{+}& C_F^2  \, n_f^2
\left[
\frac{5713}{6912} 
-\frac{581}{96}  \,\zeta_{3}
+\frac{3}{4}  \,\zeta_3^2
+\frac{125}{24}  \,\zeta_{5}
\right]
+\dots
\label{d4}
\end{eqnarray}
\begin{eqnarray}
&&\widetilde{d}_4 =  
\,C_F   \, n_f^3
\left[
-\frac{520771}{746496} 
+\frac{65}{576}  \,\zeta_{3}
+\frac{1}{192}  \,\zeta_{4}
+\frac{5}{24}  \,\zeta_{5}
\right]
\nonumber\\
&{+}& C_F^2  \, n_f^2
\left[
\frac{7009861}{497664} 
-\frac{4073}{384}  \,\zeta_{3}
\right.
\nonumber\\
&&\hspace{2.3cm}
\left.
+\frac{3}{4}  \,\zeta_3^2
-\frac{15}{128}  \,\zeta_{4}
+\frac{89}{24}  \,\zeta_{5}
\right]
\nonumber\\
&{+}& \,C_F \,C_A   \, n_f^2
\left[
\frac{18248293}{995328} 
-\frac{1243}{576}  \,\zeta_{3}
\right.
\nonumber\\
&&\hspace{1.7cm}
\left.
-\frac{1}{8}  \,\zeta_3^2
+\frac{15}{128}  \,\zeta_{4}
-\frac{2621}{576}  \,\zeta_{5}
\right]
{}+\dots ,
\label{ds4_tilde}
\end{eqnarray}
\begin{eqnarray}
 {} r_4 &=&   
 \, n_f^3
\left[
-\frac{6131}{5832} 
+\frac{11}{432}  \pi^2
+\frac{203}{324}  \,\zeta_{3}
-\frac{1}{54}  \pi^2 \,\zeta_{3}
+\frac{5}{18}  \,\zeta_{5}
\right]
\nonumber\\
&{+}& \, n_f^2
\left[
\frac{1045381}{15552} 
-\frac{593}{432}  \pi^2
-\frac{40655}{864}  \,\zeta_{3}
\right.
\label{r4_qcd}
\\
&{}& 
\hspace{1.6cm}
+\ \left.
\frac{11}{12}  \pi^2 \,\zeta_{3}
+\frac{5}{6}  \,\zeta_3^2
-\frac{260}{27}  \,\zeta_{5}
\right]
+ \dots
{},
\nonumber
\end{eqnarray}
\begin{eqnarray}
\widetilde{r}_4 &=&   
 n_f^3
\left[
-\frac{520771}{559872} 
+\frac{275}{3888}  \pi^2
-\frac{1}{2160}  \pi^4
\right.
\label{rs4_qcd}
\\
&& 
\hspace{0.8cm}
+
\left.
\frac{65}{432}  \,\zeta_{3}
-\frac{1}{54}  \pi^2 \,\zeta_{3}
+\frac{1}{144}  \,\zeta_{4}
+\frac{5}{18}  \,\zeta_{5}
\right]
\nonumber\\
&{+}& \, n_f^2
\left[
\frac{220313525}{2239488} 
-\frac{197119}{31104}  \pi^2
-\frac{11875}{432}  \,\zeta_{3}
+\frac{5}{6}  \,\zeta_3^2
\right.
\nonumber\\
&{+}& 
\left.
\frac{143}{4320}  \pi^4
+\frac{259}{216}  \pi^2 \,\zeta_{3}
+\frac{25}{96}  \,\zeta_{4}
-\frac{5015}{432}  \,\zeta_{5}
\right]
+ \dots
\nonumber
\end{eqnarray}
In (\ref{d4})  and (\ref{ds4_tilde})   we have 
distinguished between contributions proportional to
different color factors, while  for (\ref{r4_qcd}) and (\ref{rs4_qcd}) 
their  SU(3) values  $C_F = 4/3$, $C_A = 3$
have been specified.
In (\ref{r4_qcd})  and (\ref{rs4_qcd})
we differentiate  between $\zeta_4$-terms  and 
$\pi^4$-ones generated by  the analytic continuation 
from spacelike to timelike regions.
At last, numerically $R$ and $\widetilde{R}$ read
\bea
&&
R(s) = 
1 + a_s +
 a_s^2 \left( 1.986 - 0.1153 \ n_f
\right)
\nonumber
\\
&+& a_s^3
\left(
-6.6369 - 1.2001 \ n_f - 0.00518 \ n_f^2   
\right)
\\
\nonumber
&+& a_s^4
\left(
 0.02152 \ n_f^3 - 0.7974 \ n_f^2  + \ n_f \ r_{(4,1)} + r_{(4,0)} 
\right)
{},
\label{RV:num}
\eea
\bea
&& 
\widetilde{R}(s) = 
\nonumber
1 + 5.6667 \ a_s +
 a_s^2 \left(35.940- 1.3586 \ n_f
\right)
\\
&+& a_s^3
\left(164.1392 - 25.7712 \ n_f + 0.2590 \ n_f^2
\right)
\\
\nonumber
&+& a_s^4
\left(
- 0.02046 \ n_f^3 + 9.6848 \ n_f^2  + \ n_f \ \tilde{r}_{(4,1)} + \tilde{r}_{(4,0)} 
\right)
{}.
\label{RS:num}
\eea

The result is easily applied to the semileptonic $\tau$ decay rate.
Numerically one finds for the QCD corrections to the parton result
in the massless limit 
\bea
R^{\tau} = 1 &+&   a_s + 5.2023 \ a_s^2 +  26.3659 \ a_s^3
\nonumber
\\
&+& 
 a_s^4 \
(6.1827 n_f^2  - 0.06368 n_f^3 ) 
\label{}
{},
\eea
where we have set $\mu = M_\tau$ and $n_f =3$ everywhere except for
the ${\cal O}(\as^4)$ term.

Another  application of our results is the QED
$\beta$-function. At four-loops it is  known from \cite{GorKatLarSur91}.
At  five-loops the only available information is
the leading at large $n_f$    term of order $\alpha^6 n_f^4$ 
\cite{QED-beta:renorm.chain}. For QED with $n_f$ electron flavours we find 
a partial \footnote{Partially
because  in this order the QED $\beta$-function receives  contributions
also from  so-called singlet 
diagrams, see Fig.~1b. 
  }  
result for the term of order ${\cal O}(\alpha^6 n_f^3)$, viz.
\begin{eqnarray}
&&\hspace{-5mm} \beta_{QED}(\alpha) =   
\,n_f 
\left[
\frac{4\ A^2}{3}  
\right]
{+}  
4\,  n_f A^3
-  A^4
\left[
2  \,n_f 
+\frac{44}{9}  \, n_f^2
\right]
\nonumber\\
&{+}&  A^5
\left[
-46  \,n_f 
+\frac{760}{27}  \, n_f^2
-\frac{832}{9}  \,\zeta_{3} \, n_f^2
-\frac{1232}{243}  \, n_f^3
\right]
\nonumber\\
&{+}&  A^6
\left[
\frac{9922}{81}  \, n_f^3
-\frac{7616}{27}  \,\zeta_{3} \, n_f^3
+\frac{352}{3}  \,\zeta_{4} \, n_f^3
\right.
\nonumber\\
&{}&
\hspace{2cm}
\left.
\ +\ \frac{856}{243}  \, n_f^4
+\frac{128}{27}  \,\zeta_{3} \, n_f^4
\right]
{},
\label{beta:QED}
\end{eqnarray}
where $A=\alpha/(4\pi), \alpha= e^2/(4\pi)$ and $e$ is  the electron
electric charge. 
\begin{table}[h]
    \caption{\label{tab:FACPMS}Comparison
        of the results obtained in this
        paper with estimates based of 
       FAC, PMS, NNA and two types of Pad\'e approximation methods.
             }
\begin{ruledtabular}
\begin{tabular*}{\hsize}{l@{\extracolsep{0ptplus1fil}}c@{\extracolsep{0ptplus1fil}}r}
          $r_{(4,2)}$
        & 
          $\tilde{r}_{(4,2)}$
        & 
          \\
\colrule
        \hline
          -0.797
          &
         9.685           
          &
          exact
          \\
        \hline
         -1.06
          &
         ---
          &
       \cite{KS}(PMS), \cite{Samuel:1995jc} (PAM)
       \\
       ---
        &
      7.71
        &
       \cite{CheKniSir97} (PMS, FAC)
       \\
       0.74
        &
       25.2
        &
       \cite{HPade} (APAM)
       \\
      ---
        &
     1.00
        &
\cite{Broadhurst:2001yc} (NNA)
       \\
\end{tabular*}
\end{ruledtabular}
\end{table}

In Table I our results are compared with predictions obtained with the
help of various optimization procedures.  To get these numbers we have
taken the predictions of
\cite{KS,Samuel:1995jc,CheKniSir97,HPade,Broadhurst:2001yc} (where
available) for $n_f =3,4,5$, have subtracted the ${\cal O}(\als^4
n_f^3)$ contribution and fitted the result against the quadratic
polynomial in $n_f$.  For both correlators the predictions based on
the Principle of Minimal Sensitivity (PMS) or on fastest apparent
convergence (FAC) or, at last, on the  Pad\'e  Approximation Method (PAM)
are in quite good agreement to the exact results. On the other hand,
Asymptotic Pad\'e-Approximant Method (APAM) as well as Naive
NonAbelization (NNA) fail to predict even the rough size of the
$\als^4n_f^2$ coefficients.  All in all one may conclude that
there is a good chance that
\begin{itemize}
\item  
the  often used   prediction by  Kataev  and Starshenko for the 
complete  ${\cal O}(\als^4)$ contribution to the vector correlator   will 
be  close to  reality;
\item  
 the relatively worser apparent convergence of the 
perturbative series for the scalar correlator as revealed by 
the calculation of Ref.~\cite{gssq}  will   persist in higher orders
in agreement with FAC/PMS estimations of Ref.~\cite{CheKniSir97}.
\end{itemize}

Finally, we note that the present calculation does not include all
possible topological types of diagrams appearing in order $\als^4$.
Nevertheless, our experience shows that these new topologies can be
solved in the same way, given  sufficient computer
resources. Work in this direction is  in progress.

\begin{acknowledgments}
The authors are grateful to A.~A.~Pivovarov for a useful discussion.
This work was supported by the DFG-Forschergruppe
{\it ``Quantenfeldtheorie, Computeralgebra und Monte-Carlo-Simulation''} 
(contract FOR 264/2-1), by  INTAS (grant
00-00313), by  RFBR (grant 01-02-16171) and by
the European Union under contract
HPRN-CT-2000-00149.
\end{acknowledgments}
\sloppy
\raggedright


\begin{thebibliography}{9999} 


\bibitem{CheKatTka79+NPB80} K.~G.~Chetyrkin, A.~L.~Kataev and
 F.~V.~Tkachov, Phys. Lett.  B \textbf{85}, 277 (1979);
 Nucl. Phys. B \textbf{174}, 45 (1980)
 


\bibitem{ckk96} 
K.~G.~Chetyrkin, J.~H.~K\"uhn, and A.~Kwiatkowski,  Phys. Reports. \textbf{277}, 189 (1996).


\bibitem{TESLA} TESLA, Technical Design Report, DESY 2001-011; ECFA
2001-209; Part III, {\em Physics at an $e^+e^-$ Li\-near Collider},
R.-D. Heuer, D. Miller, F. Richard, P. Zerwas, eds.



                  
\bibitem{GorKatLar91SurSam91}
S.~G.~Gorishny, A.~L.~Kataev and S.~A.~Larin, Phys. Lett.  B \textbf{259}, 144 (1991);
L.~R.~Surguladze and M.~A.~Samuel,  Phys. Rev. Lett. \textbf{66}, 560 (1991).


\bibitem{KniKue90b} B.~A.~Kniehl, J.~H.~K\"uhn, Phys. Lett.  B \textbf{224}, 229 (1990);
 Nucl. Phys. B \textbf{329}, 547 (1990).


\bibitem{QCDreview}
S.~Bethke,
J.~\ Phys.\ G {\bf G26}, R27 (2000)
J. Phys. G \textbf{G26}, R27 (2000q).

\bibitem{ste}
P.~M.~Stevenson, 
Phys. Rev. D  \textbf{23}, 2916 (1981); 
Phys. Lett.  B \textbf{100}, 61 (1981);
Nucl. Phys. B \textbf{203}, 472 (1982); 
Phys. Lett.  B \textbf{231}, 65 (1984).


\bibitem{gru}
G.~ Grunberg, Phys. Lett.  B \textbf{95}, 70 (1980); 
Phys. Rev. D  \textbf{29}, 2315 (1984).

 
\bibitem{BG}  D.~J.~Broadhurst and A.~G.~Grozin, 
Phys. Rev. D  \textbf{52}, 4082 (1995).   

\bibitem{me81b}
K.~G.~Chetyrkin and F.~V.~Tkachov,
Nucl. Phys. B \textbf{192}, 159 (1981).        


\bibitem{mincer2}
S.~A.~Larin, F.~V.~Tkachov, J.~A.M.~Vermaseren,
{ Preprint NIKHEF-H}/91-18 (1991).


\bibitem{Ver91}
J.~A.~M.~Vermaseren, \textit{ Symbolic Manipulation 
with FORM, Version 2, CAN,  Amsterdam}, 1991.  
 

\bibitem{me84} 
K.~G.~Chetyrkin and V.~A.~Smirnov,  
Phys. Lett.  B \textbf{144}, 419 (1984).

\bibitem{gvvq}
K.~G.~Chetyrkin, Phys. Lett.  B \textbf{391}, 402 (1997).
                                                      

\bibitem{VV:renormalons}
M.~Beneke, Nucl. Phys. B \textbf{405}, 424 (1993). 


\bibitem{bai1} %
P.~A.~Baikov, Phys. Lett.  B \textbf{385}, 404 (1996);
Nucl. Inst. Meth. \textbf{A389}, 347 (1997);\\
P.~A.~Baikov and  M.~Steinhauser, Comp. Phys. Commun.  \textbf{115}, 161 (1998).


\bibitem{bai2} 
P.~A.~Baikov, Phys. Lett.  B \textbf{474}, 385 (2000).

  

\bibitem{Tar} 
O.~V.Tarasov, Nucl. Phys. Proc. Suppl. \textbf{89}, 237 (2000).

\bibitem{Fliegner:2000uy}
D.~Fliegner, A.~Retey and J.~A.~Vermaseren,
hep-ph/0007221.


 
\bibitem{gssq}
K.~G.~Chetyrkin, Phys. Lett.  B \textbf{390}, 309 (1997).        



\bibitem{GorKatLarSur91}
S.~G.~Gorishny, A.~L.~Kataev,  S.~A.~Larin and  L~R.~Surguladze,
Phys. Lett.  B \textbf{256}, 81 (1991).
 


\bibitem{QED-beta:renorm.chain} A.~Palanques--Mestre and P.~Pascual,
 Comm.  Math.  Phys.  \textbf{95}, 277 (1984).   


\bibitem{KS} A.~L.~Kataev and V.~V.~Starshenko, 
Mod. Phys. Lett.  \textbf{A 10}, 235 (1995).

\bibitem{Samuel:1995jc}
M.~A.~Samuel, J.~R.~Ellis and M.~Karliner,
 Phys. Rev. Lett. \textbf{74}, 4380 (1995).

\bibitem{CheKniSir97}
 K.~G.~Chetyrkin, B.~A.~Kniehl and A.~Sirlin, Phys. Lett.  B \textbf{402}, 359 (1997).



\bibitem{HPade} F.~Chishtie, V.~Elias and T.~G.~Steele,
Phys. Rev. D  \textbf{59}, 105013 (1999).

\bibitem{Broadhurst:2001yc}
D.~J.~Broadhurst, A.~L.~Kataev and C.~J.~Maxwell,
Nucl. Phys. B \textbf{592}, 247 (2001).
\end{thebibliography}
\end{document}